\documentclass[12pt,preprint]{aastex}
\usepackage{psfig}

\newcommand{\lta}{$\; \buildrel < \over \sim \;$}
\newcommand{\simlt}{\lower.5ex\hbox{\lta}}
\newcommand{\gta}{$\; \buildrel > \over \sim \;$}
\newcommand{\simgt}{\lower.5ex\hbox{\gta}}

\slugcomment{Accepted to {\it ApJ}}
\shortauthors{Howell et al.}
\shorttitle{HST/STIS Observations of LL And and EF Peg}

\begin{document}

\title{HST/STIS Spectroscopy of the White Dwarfs \\ in the \\ Short-Period Dwarf
Novae LL And and EF Peg\footnote{Based on observations made with the NASA/ESA
Hubble Space Telescope obtained at the Space Telescope Science Institute
which is operated by the AURA under NASA contract NAS 5-26555 and with
the Apache Point Observatory (APO) 3.5-m telescope which is operated by the
Astrophysical Research Consortium (ARC).}}

\author{Steve B. Howell}
\affil{Astrophysics Group, Planetary Science Institute, Tucson, AZ  85705}
\email{howell@psi.edu}
 
\author{Boris T. G\"ansicke}
\affil{Department of Physics and Astronomy, 
University of Southampton, Southampton, SO17 1BJ, UK}
\email{btg@astro.soton.ac.uk}

\author{Paula Szkody}
\affil{Department of Astronomy, University of Washington, Seattle, WA 98195}
\email{szkody@astro.washington.edu}

\author{Edward M. Sion}
\affil{Villanova University, Villanova, PA 19085}
\email{edward.sion@villanova.edu}

\begin{abstract}
We present new HST/STIS observations of the short-period dwarf novae 
LL And and EF Peg during deep quiescence. 
We fit stellar models to the UV spectra and use
optical and IR observations to determine the physical parameters of the white
dwarfs in the systems, the distances to the binaries, and the properties of the
secondary stars.
Both white dwarfs are relatively cool, having T$_{eff}$ near 15000K, and
consistent with a mass of 0.6M$_{\odot}$. 
The white dwarf in LL And appears to be of
solar abundance or slightly lower 
while that in EF Peg is near 0.1-0.3 solar. 
LL And is found to be 760 pc away while EF Peg is closer at 380 pc.
EF Peg appears to 
have an $\sim$M5V secondary star, consistent with that expected for its orbital 
period, while the secondary object in LL And remains a mystery. 
\end{abstract}


\section{Introduction}

Observations of the component stars within interacting binaries such as
cataclysmic variables (CVs) are difficult as the collected flux is often
contaminated, representing contributions from both of the component stars as
well as accretion phenomena. We have undertaken a large HST program to 
obtain UV spectroscopy of the white dwarfs in a number of short-period dwarf
novae. Our selection of short period CVs was made primarily on the basis of
systems with low mass accretion rates (based on short-orbital period and 
outburst history) and actually detecting the white dwarf absorption
wings in the optical spectra. Thus, UV spectroscopy was essentially
guaranteed to provide data in which the white dwarf would contribute 
$\sim$90\% of the UV continuum flux,
allowing accurate modeling of its parameters.
In addition to our ability to fit white dwarf models to the UV spectra, 
we can use the
observations to set limits on the distance to the binary and as a fiducial
point to construct 
multi-wavelength spectral energy distributions to be attempted.

In this paper, we report on our analysis of Hubble Space Telescope (HST)/Space
Telescope Imaging Spectrograph (STIS) observations for the two
short-period dwarf novae: LL And and EF Peg. LL And belongs to the
class of short-period dwarf novae which have rare outbursts, only of the
``super" variety, increasing in brightness by 6 magnitudes or more.
Additionally, both systems have been suspected as CVs which may 
harbor very low mass,
brown dwarf-like, secondary stars and be of extreme age. These properties 
have been used to define a small class of objects as Tremendous Outburst
Amplitude Dwarf novae (TOADs; Howell, Szkody \& Cannizzo 1995) 
of which the most famous member is WZ Sge. EF Peg is suspected
as being a TOAD due to its large outburst amplitude. 
However, its inter-superoutburst time scale is not well known and may be
as short as $\la$1 year and normal outbursts may also occur 
(Howell et al. 1993). If these outbursts characteristics are 
true, EF Peg is unlikely to be a TOAD.

LL And has an orbital period near 78 minutes based on superoutburst
observations (Howell and Hurst 1994) with no radial velocity solution yet
available. Optically, LL And is currently in a deep quiescent state with 
V$\sim$20
and with its last superoutburst being in 1993. Szkody et al. (2000) used an
optical spectrum, which revealed the underlying white dwarf, and fit an 11,000K,
0.6 M$_{\odot}$ white dwarf model. Howell and Ciardi (2001) presented K-band 
spectroscopy of LL And which showed methane absorption 
indicating a very low effective temperature near 1300K.
However, since Szkody et al. (2000) determined a lower limit to the 
distance for LL And of 346 pc, it is difficult to reconcile the methane
absorption as originating in a low mass brown dwarf-like secondary
star.

EF Peg has a longer orbital period, 2.05 hours, based on superoutburst 
photometry
(Howell et al. 1993) but has been poorly studied due to its
faintness, V$\la$18.5 at minimum and its positional location within 5"
of a 12th magnitude star, long thought to be the variable (See Howell et al.
1993). EF Peg had its last superoutburst in 1997 and no previous 
optical spectrum for EF Peg exists. 
Howell et al. also present a crude distance
estimate for EF Peg of $\sim$600 pc based on a normal dwarf novae
M$_V$--P$_{orb}$ relation (Warner 1995). 
If a TOAD, EF Peg would be the longest orbital period member and therefore
may be one of the oldest CVs in the Galaxy (Howell, Rappaport and Politano
1997). Recent K-band spectroscopy of EF Peg has been
presented in Littlefair et al. (1999) and these authors concluded that the
secondary star must be later than (cooler than) $\sim$M5V. 

Details of our motivations, the HST program in general, and previous results
are presented in
G\"ansicke et al. (2001) and Szkody et al. (2002). Our new HST observations are
discussed in \S2, our model fitting and analysis is given in \S3, 
and we summarize the results of this work in Section 4.

\section{Observations}

HST/STIS observations of LL And and EF Peg were obtained on 
12 July 2000 UT and 18 June 2000 UT using the G140L and E140M gratings
respectively
with the 52 X 0.2" aperture. The UV spectra covered the range of 
1150-1715 \AA~with an estimated velocity resolution of 
$\sim$300 km s$^{-1}$. The total exposure
times were 75 min for LL And (2 consecutive HST orbits) and 115 min for EF Peg 
(3 consecutive HST orbits). 
The summed
spectra covered 93\% (LL And) and 95\% (EF Peg) of their orbital periods
but were not consecutive in phase coverage due to the sampling
imparted by the HST observations. Since neither star has a radial velocity
solution and thus no ephemeris, the binary phases of the individual HST 
observations are unknown. We will see below, that only the summed observations
had sufficient S/N to use for our purposes.
The mean HST/STIS spectra of LL And and 
EF Peg are shown in Fig\,1a and 1b, respectively

The UV spectrum of LL And (Fig. 1a) shows the typical disk emission
line at CIV (1550 \AA). 
%
%
A number of weak absorption lines can be clearly discerned in the
spectrum: the CI 1277-1280\,\AA\ multiplet, CI 1335\,\AA\, and SiII
1306, 1527 \& 1530\,\AA (Fig.\,5 below). A number of additional
transitions are marginally identified near the noise level.  These
absorption lines are almost certainly from the white dwarf as they are
not significantly broadened by rotation (as the disk emission lines
are) and the observed ionization/excitation exactly matches the
derived white dwarf photospheric temperature which we derive below. It
is possible, though unlikely, that a disk at a low enough inclination
could produce these absorption lines but at the distances derived (see
below) luminosity arguments strongly support the photosphere
interpretation.  The quality of the spectrum in the wavelength range
near 1300\AA~ is degraded due to the subtraction of the Geocoronal OI
airglow. Residual Ly$\alpha$ airglow is visible in the broad white
dwarf Lyman absorption.  For EF Peg (Fig. 1b), we see emission due to
CIV as well as NV and SiIII and IV.  Narrow absorption lines of CI
(1277\AA), SiII (1260/65 \AA), and CII (1335 \AA) are also
observed. Quasi-molecular hydrogen absorption troughs (H$^+_2$ at
1400\AA~and H$_2$ at 1600\AA) are not obvious in either star being
weak at these temperatures and probably filled in by SiIV disk
emission and/or lost in the continuum noise.

The STIS data were obtained in the {\it time-tagged} mode
and we made use of this information to examine
the photometric behavior of these two stars during our HST observations.
Excluding a small region near Ly$\alpha$, we produced 10-30 sec time resolution
UV light curves by summing the continuum flux. 
Neither star showed any significant modulation during our
observations, consistent with the assumption
that nearly all the UV continuum light
is from the white dwarf and does not contain a significant accretion disk/hot
spot contribution.

Neither star has any previous UV spectra but given the rarity of their
superoutbursts (last superoutburst for LL And was in 1993 and 
for EF Peg in 1997) 
as well as the AAVSO and other watchdog groups who monitor
the TOADs for such outbursts not reporting any activity during our HST
observations, we conclude that our UV spectra were taken during deep quiescence.
Additionally, we will see below that both stars are at or below their reported
minimum V magnitudes.

Optical spectra for EF Peg (3 July 2000 UT: 
within one month of the HST observation) 
and LL And (5 December 2000 UT) were obtained 
using the Apache Point
Observatory 3.5m telescope and Double Imaging Spectrograph with a 1.5 arcsec
slit, giving $\sim$2.5\AA\ resolution spectra in the regions from
4200-5000\AA\ and 6300-7300\AA. 

LL And is a fairly faint target 
and was observed for 1200 sec at an airmass of 2.8. Light cirrus was present
during the observation, so the fluxes are highly uncertain. 
Although of low S/N (especially in the blue), the spectra 
obtained are consistent with that presented in Szkody et al. (2000) and
are shown in Figure 2a as we use them in \S3.3. The 5000\AA~flux
corresponds to an approximate V magnitude of $\sim$20.6, fainter than the V=19.9
value reported by Szkody et al. (2000) for LL And in 1996
when only 3 years past its last superoutburst. Given the high airmass
of our observation and the cirrus present, we take these values as being
consistent. 
The Szkody et al. spectrum of LL And 
showed the presence of underlying WD absorption in
H$\beta$ and the higher Balmer lines, but we cannot conclusively
detect the absorption
in our poor quality blue spectrum. 
However, the optical spectra of LL And 
do show the presence of double-peaked 
emission lines for H$\alpha$ and H$\beta$
and a weak blue continuum of a dwarf novae at minimum. 
Both of these properties are indications of a system with low mass transfer and 
a mostly optically thin accretion disk (See Mason et al. 2000).
Given the rare outbursts of this star, these properties are all 
typical signs of a quiescent state TOAD.

For EF Peg, 3x900 sec exposures were obtained over an airmass range of 1.6-1.9.
The average spectrum is shown in Figure 2b and converting the flux at 
5000~\AA~to 
an approximate visual magnitude gives V$\sim$19. 
Even though the observation was through a narrow 
slit and there were passing clouds
during the night, we find that it is in 
good agreement with the published quiescent 
V magnitude for EF Peg (Howell et al. 1993).
We see clear evidence for white dwarf absorption
in EF Peg at H$\beta$ and H$\gamma$. 
Figure 2b also 
shows the typical double-peaked Balmer 
line profiles and weak blue continuum of a low mass transfer rate
CV at minimum light. 

\section{Model Fitting and Analysis}

Our white dwarf spectral analysis is carried out in two stages: in a
first step, we derive the effective temperature $T_\mathrm{eff}$ and
the surface gravity $\log g$ as well as a rough estimate of the
chemical abundances in the atmosphere. In a second step, we refine the
analysis of the abundances and determine the white dwarf rotation
rate. All the model spectra used throughout this analysis are
generated using TLUSTY195 and SYNSPEC45 (Hubeny 1988; Hubeny \& Lanz
1995).

During the first stage, we calculate three different grids of model
spectra covering $T_\mathrm{eff}=10,000-25,000$\,K in steps of 100K
and $\log g=7.0-10.0$ in steps of 0.1, assuming 0.1, 0.5, and 1.0
times solar abundances. We fit the models to the HST/STIS data with
the emission lines (presumably originating in the accretion disk)
added in as simple Gaussian profiles in the fitting procedure and we
exclude from the fit a 20\AA~wide region centered on Ly$\alpha$. 

In principle, one can determine from such model fits both the
temperature and the $\log g$ independently as both parameters
determine the detailed shape of the continuum and line
profiles. However, in the case of LL\,And and EF\,Peg, this is
difficult as the data have a relatively modest S/N and the Ly$\alpha$
absorption profile is contaminated with both disk and terrestrial
emission, especially in the case of EF\,Peg. In order to test the
correlation between $T_\mathrm{eff}$ and $\log g$, we keep $\log g$
fixed and determine the best-fit effective temperature, repeating this
procedure for each individual $\log g$ in our model grid. We find that
$T_\mathrm{eff}$ and $\log g$ are approximately linearly correlated,
with statistically insignificant differences in $\chi^2$.

Using a Hamada-Salpeter mass-radius relation (Hamada \& Salpeter 1961)
for carbon-core white
dwarfs\footnote{The use of the zero-temperature $M-R$ relation is
justified for our purpose, as the increase in radius of
finite-temperature white dwarfs is small in the temperature range
considered here, and insignificant compared to other uncertainties.}
we compute from $\log g$ the white dwarf mass $M_\mathrm{WD}$ and its
radius $R_\mathrm{WD}$. From the scaling factor (R$_{WD}^2$/d$^2$) 
between the model
spectrum and the observations (assuming negligible extinction for
these high-galactic latitude objects), we obtain an estimate for the
distances of the systems. Finally, we compute $V$ magnitudes from the
optical extensions of the model spectra used for the STIS
analysis. Using our initial white dwarf model grid, we calculated $\chi^2$
values (over the entire range of $\log g$) for both stars. For LL And, a
minimum from $\log g$=7.2-8.1 was found with $\chi^2$ $\la$1.1. For EF Peg, 
$\chi^2$ $\la$2.8 over the minimum covering $\log g$=7.0-8.2.
The parameters $M_\mathrm{WD}$, $T_\mathrm{eff}$, $d$, and $V$ are shown as
a functions of $\log g$ in Fig.\,4 (LL\,And) and Fig.\,7
(EF\,Peg). Over a fiducial range $\log g=8.0\pm0.5$ (corresponding to
$M_\mathrm{WD}=0.33-0.90\,M_{\odot}$), we find $T_\mathrm{eff}$ to be
constrained to $\pm1000$\,K.

While the STIS data alone are insufficient to determine independently
the temperature and the mass of the white dwarf, additional
constraints can be obtained from combining the UV data with optical/IR
coverage. The distance value determined 
critically depends on the assumed white dwarf radius but the use of equal
(minimum) state, optical (and IR) observations help constrain it as well. 
Using our models over a range of $\log g$, we can assess their temperature
dependence and thus their flux contribution to longer wavelength bandpasses.
A lower limit to $\log g$ can be obtained by assuming that the optical flux can
not lie below our white dwarf model. An absolute upper limit to $\log g$ is
established by the theoretical upper mass limit (1.4 M$_{\odot}$) 
for a white dwarf.
We discuss these ideas further in \S 3.3 but note here that, when taken
together, all the observed parameters favor a best choice for $\log g$=8.0 for
both stars.


After having established ($\log g$, $T_\mathrm{eff}$) 
from the combined ultraviolet plus optical
observations, we made the attempt to refine the models in terms of the
chemical abundances and white dwarf rotation rates.  We plotted our STIS
data along with model spectra computed for 0.1, 0.5, and 1.0 times solar
abundances, convolving each of these models with rotation rates of
$v\sin i=100-1200\,\mathrm{km\,s^{-1}}$. 
Considering the noise level of the data, we judged the
goodness of the match between data and model by visual inspection rather
than by a formal fit. Our experience is that at times, such formal fitting to 
data of this S/N can lead to false estimations of the best $\chi^2$ fit.

\subsection{LL And}

Our HST spectrum of LL\,And is rather noisy due to the intrinsic
faintness of this object and the short integration time, thus we
co-added both HST orbits for our analysis. Using 
$\log g=8.0\pm0.5$, which corresponds to 
a white dwarf mass of $0.6\pm0.3 M_{\odot}$, we find an effective temperature
of $14,300\pm1000$\,K and a distance of $760\pm100$\,pc for LL And. 
This model is
shown in Figure 3 and the dependence of our determined parameters on $\log g$ is
presented in Figure 4. 
These model
parameters predict a white dwarf only apparent magnitude of
V=20.8, in good agreement with our minimum system $V$ magnitude of near
20.6. Thus, we expect that at the present time the accretion disk 
contributes little flux to the (blue) optical bandpass; a result which
agrees with our observation of no significant blue continuum rise
(Figure 2a).

The moderate S/N of the STIS data prevents a detailed analysis of the
abundance pattern, but our fits {\it suggest} slightly sub-solar
abundances for the atmosphere of the white dwarf. While there are
only a very limited number of metal absorption lines available that
can be used to measure the rotation rate, we are confident
that the white dwarf spin is significantly below Keplerian velocity;
we estimate $v\sin i\la500\,\mathrm{km\,s^{-1}}$ (see Fig.\,5) where it can be
seen that a best match occurs for velocities below 500 $\mathrm{km\,s^{-1}}$.

\subsection{EF Peg}

Due to the moderate S/N in each of the three HST orbits of STIS
spectra for EF Peg, we have summed the data together for our analysis
presented here.  
Our model for EF Peg
is shown in Figure 6 and the dependence of 
our determined parameters on $\log g$ is
presented in Figure 7. 
We find a white dwarf mass of 0.6$\pm$0.3 M$_{\odot}$ ($\log g$=8.0$\pm0.5$).
a WD temperature of
16,600$\pm$1000K, a distance of 380$\pm$60 pc, 
and a white dwarf only V magnitude of
19.1.  Our ``spectral" V magnitude of $\sim$19 again assures us of
little accretion disk flux contamination in the (blue) optical bandpass,
as seen in Fig. 2b as a lack of a rising blue continuum.

The determination of the elemental abundances present in EF Peg is
{\it uncertain} due to the (unknown amount of) contamination of line
emission in and near the absorption features, especially for the
carbon and the silicon lines. The overall abundances appear to be
sub-solar ($\sim0.1-0.3\times$ solar).
We have used these abundance estimates to
provide an adequate fit (Fig. 6) to the UV spectra. 
Figure\,8 shows the STIS data of
EF\,Peg along with the white dwarf model just described convolved with
rotation rates of $v$sin$i$=100, 300, and 500
km\,s$^{-1}$. Examination of the different absorption lines, in
particular  the SiII 1260/65\,\AA\ doublet, shows that the rotation rate
of the white dwarf in EF\,Peg is very low, the best match is indeed
obtained for $v$sin$i$$\la$300 km\,s$^{-1}$.

\subsection{Spectral Energy Distributions and the Secondary Stars}

In order to provide an additional constraint on the $\log g$ 
values determined in
the UV fitting process, we can make use of the optical spectra obtained for 
LL And and EF Peg. Using the assumption that the ``state" of the system was
similar, the optical spectra set a limit on the WD flux as it cannot be higher
than the value measured with any additional optical flux being due to
accretion phenomena. We have also seen above that for both of these stars
during the time of the optical observations, 
there is likely to be little contamination in the
optical, particularly in the blue region, due to the accretion disk.

Figure 9 shows the spectral energy distribution (SED) for LL And based on our
UV and optical spectroscopy.
The UV models in the optical bandpass, show that the $\log g$
= 7.5 model is unlikely as it would provide too much optical flux and the $\log
g$
= 8.5 model provides too little to the blue (uncontaminated) optical flux.
Thus, a $\log g$=8.0 model is the most plausible fit for these data.

In Fig. 9, we also plot the K band magnitude 
obtained by Howell \& Ciardi (2001) for LL And based on IR spectroscopy.
Their low S/N spectrum revealed the presence of 
methane absorption in LL And
and, if confirmed, would make the secondary star quite cool (T$\sim$1300K) and
of very low mass ($\sim$0.04 M$_{\odot}$), in agreement with theoretical
values if a post-period minimum CV. However, for such a secondary star
the distance to LL And would need to be quite close, $\sim$30-60 pc, a result
in severe disagreement with our value of 760 pc determined here and the lower
limit of 346 pc determined by Szkody et al. (2000). The methane feature
may possibly arise in the cool outer regions of the accretion disk while
the true nature of the secondary star in LL And remains a mystery.

Figure 10 shows similar SEDs for EF Peg. The details of the models 
again provide a good constraint for our choice of a $\log g$=8.0 model 
as lower $\log g$ models would provide too much optical flux.
In Fig. 10 we show the J and K magnitudes for EF Peg from Sproats et al.
(1996). A check of the AAVSO records for EF Peg during the time of
these observations (September 1993) reveals that the star was not in
(super)outburst at the time of the IR observations. 

Using the orbital period of EF Peg as a guide,
we can estimate the expected parameters of its secondary star under the
assumptions of it being either a pre- or a post-period minimum dwarf novae
(Howell et al. 2001). For a pre-period minimum system, the secondary star
should have T$\sim$2900K and a mass of $\sim$0.2 M$_{\odot}$, similar to
the properties of a M5V star. Using the two observed IR data points as an
absolute scaling for a model fit (assuming zero disk contribution in the
optical and IR) and taking 
a SED for a main sequence star, we can add into
Fig. 10 the expected flux values at 6400\AA~and 7000\AA~ and the expected
SED shape (dashed line) for a Roche-lobe filling M5V star.
We find that for this type of secondary, we would not expect to detect it
in our red optical spectrum, as its flux would be far less than that 
of even the WD alone at 7000\AA. Assuming that the IR flux is almost entirely 
due to this type of a secondary star, a distance of $\ga$290$\pm$50 pc 
is derived. This independent
distance estimate is in good agreement with our determination above 
(380$\pm$60 pc) based on our assumed WD models.

The IR
points only provide an upper limit as the actual percentage 
of secondary star and disk
contribution to the continuum longward of 7000\AA~is unknown. 
However, the secondary star is unlikely to be too much later than 
$\sim$M5V in order for the
IR continuum shape to agree with observations.
We note that a post-period minimum secondary star of extreme low
temperature and mass would not be able to provide the observed IR fluxes unless
the system was very nearby {\it or} 
the accretion disk dominates in the IR and has a
spectral shape similar to an M star. However, both of these 
facts would make the WD 
contribution in the UV and optical very difficult to reconcile. 
Assuming the secondary star is similar to an M5V, EF Peg is not likely to be a
TOAD.

Subtraction of the WD (and secondary star for EF Peg) 
models from our optical spectra
reveal that for LL And there is not much remaining optical flux. This result
is consistent with a system of very low mass transfer and containing an
optically thin accretion disk (See Mason et al. 2000). For EF Peg, we find
residual flux remains in the red spectrum with a poorly constrained slope of
F$_{\lambda}$$\sim$$\lambda^{-3}$, possibly indicating that the outer disk is
marginally optically thick or that there is a weak single blackbody-like source
(hot spot?) contributing in this region (See Mason et. al. 2001).

\section{Summary}

Using white dwarf models, our HST
UV spectra of the stars LL And and EF Peg,
and the additional constraints imposed by optical and IR measurements,
we find the following: 
the white dwarfs in both of these short-period dwarf novae are consistent with
a mass of 0.6M$_{\odot}$($\log g$= 8.0) and show derived 
temperatures near 15,000K.
Using the most plausible models, 
we estimate the distance to LL And as 760 pc and to EF Peg as
380 pc. This distance estimate for EF Peg is considerably less than that 
calculated by Howell et al. (1993) using ``normal" dwarf novae ideas
of the absolute magnitude of the accretion disk at maximum (Warner 1995).
The failing of that method is not unexpected, as we have ample evidence
that the accretion disks in low mass transfer dwarf novae are not ``normal" at
minimum, thus they probably do not follow a normal M$_V$ relation at maximum.
UV spectroscopy is a very important technique for providing good distance
estimates for faint, short period cataclysmic variables as the usual
null detection of the secondary star can, at best, only provide a limit.
The abundance determinations from this work are not well constrained
but indicate that the white dwarf in LL And is near or slightly below solar
while that in EF Peg is near 0.1-0.3 solar.

Given the orbital periods of LL And and EF Peg, our rotation rate estimates for
the white dwarfs in these two stars are generally consistent with the 
range evident in short period sytems (Sion 1999, Gansicke et al. 2001, Szkody
et al. 2002). The quality of the
LL And UV spectroscopy precludes us from setting a strict limit.



If the duty cycle for short-period, low mass transfer rate dwarf novae in
deep quiescence is long,
observational searches for very faint (V$>$18), essentially
cool white dwarfs with (weak) Balmer emission 
and {\it close}, very
low mass (substellar?) companions in extremely short orbital periods will be a
difficult project. 
IR observations will be extremely helpful, however
current surveys, such as 2MASS, are not yet deep
enough to provide JHK measurements for the faintest CVs and their
brown dwarf-like secondaries.
Such a challenging observational program may be
required to truly test the existence of the large 
hypothesized population of TOADs.

The authors would like to thank the anonymous referee for their comments.
This research was partially supported by NASA through grant GO-08103-97A
from the Space Telescope Science Institute, which is operated
by the Association of Universities for Research in Astronomy, Inc. 
under NASA contract NAS5-26555. Additional support was provided by NSF
grant AST-98-10770 to SBH and NSF grant AST-99-01955 to ES. BTG is supported
by a PPARC Advanced Fellowship.

\references

\reference{} G\"ansicke, B. T., Szkody, P., Sion, E. M., Hoard, D., Howell, S.
B., Cheng, F. H., \& Hubeny, I., 2001, A\&A, 374, 656

\reference{}  Hamada, T. Salpeter, E. E., 1961, ApJ 134, 683

\reference{} Howell, S. B., Schmidt, R., DeYoung, J. A., Fried, R., Schmeer,
P., \& Gritz, L., 1993, PASP, 105, 579

\reference{} Howell, S. B., \& Hurst, G., 1994, JBAA, 106, No. 1, p. 29

\reference{} Howell, S. B., Szkody, P., \& Cannizzo, J., 1995, ApJ, 439, 337

\reference{} Howell, S. B., Rappaport, S. \& Politano, M., 1997, MNRAS, 287,
929

\reference{}  Howell, S. B., Nelson, L., \& Rappaport, S, 2001, ApJ, 550, 518

\reference{} Howell, S. B., \& Ciardi, D. R., 2001, ApJL, 550, L59

\reference{} Littlefair, S. P., Dhillon, V. S., Howell, S. B., \& Ciardi, D.
R., 1999, MNRAS, 313, 117

\reference{}
Mason, E., Skidmore, W., Howell, S. B., \& Mennickent, R., 2001, ApJ, 563, 361

\reference{} Mason, E., Skidmore, W., Howell, S. B., Ciardi, D., Littlefair, S.,
\& Dhillon, V., 2000 MNRAS, 318, 440

\reference{} Sion, E., M., Long, K. S., Szkody, P., \& Huang, M., 1994, ApJL, 
430, L53

\reference{} Sproats, L. N., Howell, S. B. and Mason, K. O., 1996, MNRAS, 282,
1211

\reference{} Szkody, P., Desal, V., Burdullis, T., Hoard, D., Fried, R.,
Garnavich, P., \& G\"ansicke, B., 2000, ApJ, 540, 983

\reference{} Szkody, P., G\"ansicke, B., Sion, E.,\& Howell, S. B., 2002,
in {\it The Physics
of Cataclysmic Variables and Related Objects}, B. G\"ansicke, K. Beuermann,
and K. Reinsch Eds., ASP Conf. Series, 261, 21


\reference{} Warner, B., 1995, {\it Cataclysmic Variable Stars}, Cambridge
University Press 

\newpage

\begin{deluxetable}{lccccccc}
\tablewidth{6.5in}
\tablenum{1}
\tablecaption{Summary of White Dwarf Model Fits}
\tablehead{
\colhead{Star} &
\colhead{T$_{eff}$} &
\colhead{$\log g$} &
\colhead{Mass} &
\colhead{D} &
\colhead{V$_{WD}$} &
\colhead{Abundance} &
\colhead{$v$sin$i$} \\
\colhead{} &
\colhead{(K)} &
\colhead{(cm s$^{-2}$)} &
\colhead{(M$_{\odot}$)} &
\colhead{(pc)} &
\colhead{(Mag)} &
\colhead{$\odot$} &
\colhead{(km s$^{-1}$)} 
}
\startdata
LL And & 14300$\pm$1000 & 8.0$\pm$0.5 & 0.6$\pm$0.3 & 760$\pm$100 & 20.8 & 
$\la$1.0 &
$\la$500\\
EF Peg & 16600$\pm$1000 & 8.0$\pm$0.5 & 0.6$\pm$0.3 & 380$\pm$60 & 19.1 &
0.1-0.3 & $\la$300\\
\enddata
\end{deluxetable}{}

\newpage

\begin{center}
Figure captions
\end{center}

Figure 1: (a) Total STIS spectrum of LL And binned by 0.6\AA and
smoothed with a 3-point running boxcar. The bottom curve shows the
1$\sigma$ statistical error of the \textit{unsmoothed} data. 
Note the presence of
CIV emission and the narrow geocoronal Ly$\alpha$ emission sitting in
the broad white dwarf Ly$\alpha$ absorption. Some narrow white dwarf
absorption lines are apparent, see text for details.  (b) Total STIS
spectrum of EF Peg binned by 0.6\AA. The bottom curve shows the
1$\sigma$ statistical errors, the wavelength-dependence of the error over the
individual echelle orders can be seen in the sawtooth-like
structure. Note the presence of CII-IV, and NV emission and the narrow
geocoronal Ly$\alpha$ emission sitting in the broad white dwarf
Ly$\alpha$ absorption. Some narrow white dwarf absorption lines are
apparent, see text for details. 

Figure 2: (a) Red and blue optical spectra for LL And obtained at APO. The blue
channel is poor quality due to the faintness of the source, but the
double-peaked nature of H$\alpha$ and H$\beta$ are apparent.
The y axis is Flux in units of ergs cm$^{-2}$s$^{-1}$\AA$^{-1}$.
(b) Red and blue optical spectra for EF Peg obtained at APO. The double peaked
structure of the Balmer emission lines is apparent as well as the broader
underlying white dwarf absorption features in H$\beta$ and H$\gamma$.
The y axis is Flux in units of ergs cm$^{-2}$s$^{-1}$\AA$^{-1}$.

Figure 3: Our white dwarf model (thick line) 
compared with the LL And STIS data. The 
model has T=14,300K, M$_{WD}$=0.6 M$_{\odot}$ ($\log g$=8.0), and 
the abundances listed in Table 1. 
The Gaussians (dotted lines) represent our simple fits to the emission lines.

Figure 4: Our parameter plot for LL And showing the fitting parameters
for M$_{WD}$, temperature, distance, and white dwarf V magnitude as a function
of $\log g$.

Figure 5: Three rotation models for LL And all with $v$ sin $i$=300 km s$^{-1}$
with three different values of the abundance; 0.1, 0.3, and 1.0 times solar.
We see that while none of the abundance models fit perfectly, a solar abundance
model is reasonable. Useful photospheric absorption lines are marked.

Figure 6: Our white dwarf model (thick line) compared with the
EF Peg STIS data. The 
model has T=16,600K, M$_{WD}$=0.6 M$_{\odot}$ ($\log g$=8.0),  
and the abundances listed in Table 1.
The Gaussians (dotted lines)
represent our simple fits to the emission lines.

Figure 7: Our parameter plot for EF Peg showing the fitting parameters
for M$_{WD}$, temperature, distance, and white dwarf V magnitude as a function
of $\log g$.

Figure 8: Three rotation models for EF Peg with 
0.1 solar metal abundances. The 100 km s$^{-1}$ model 
appears to be the best fit.
Useful photospheric absorption lines are marked.

Figure 9: SED model for LL And using HST/UV and APO/optical spectra.
The three white dwarf models shown cover the range of interest. The single point
at 22000 \AA~is the K band flux from Howell \& Ciardi (2001). 
The spectra have been smoothed as we are concerned here with a broad band
comparison.

Figure 10: SED model for EF Peg using HST/UV and APO/optical spectra.
The three white dwarf models shown cover the range of interest. The points 
at 12200 and 22000 \AA~are the J and K band fluxes from Sproats et al. (1996).
The dashed curve and the points at 6400 \& 7000 \AA~are a M5V 
model SED scaled to the J and K fluxes. See text for details.
The spectra have been smoothed as we are concerned here with a broad band
comparison.

\newpage

\begin{figure}
\psfig{figure=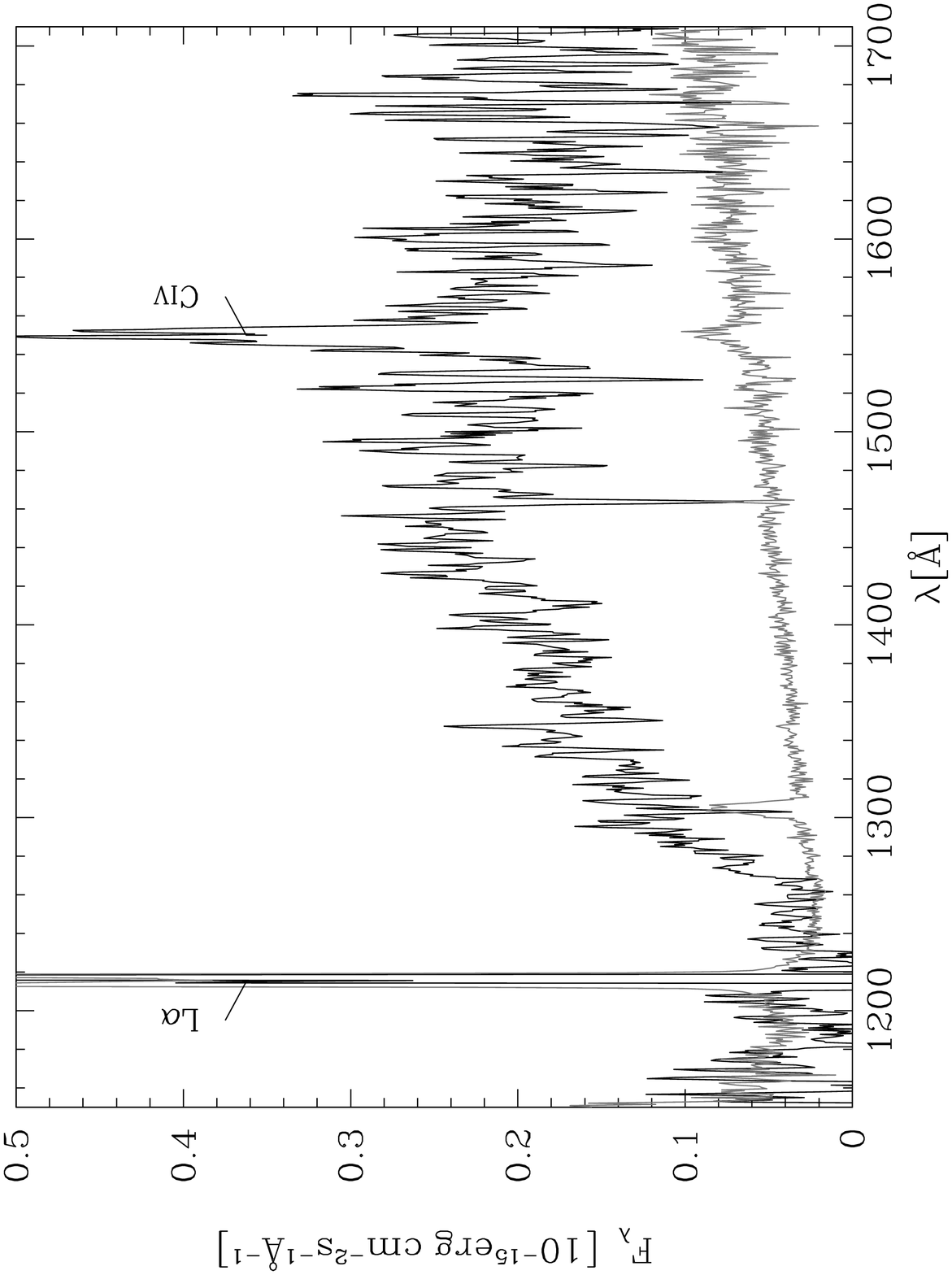,width=6in,angle=0}
\end{figure}

\begin{figure}
\psfig{figure=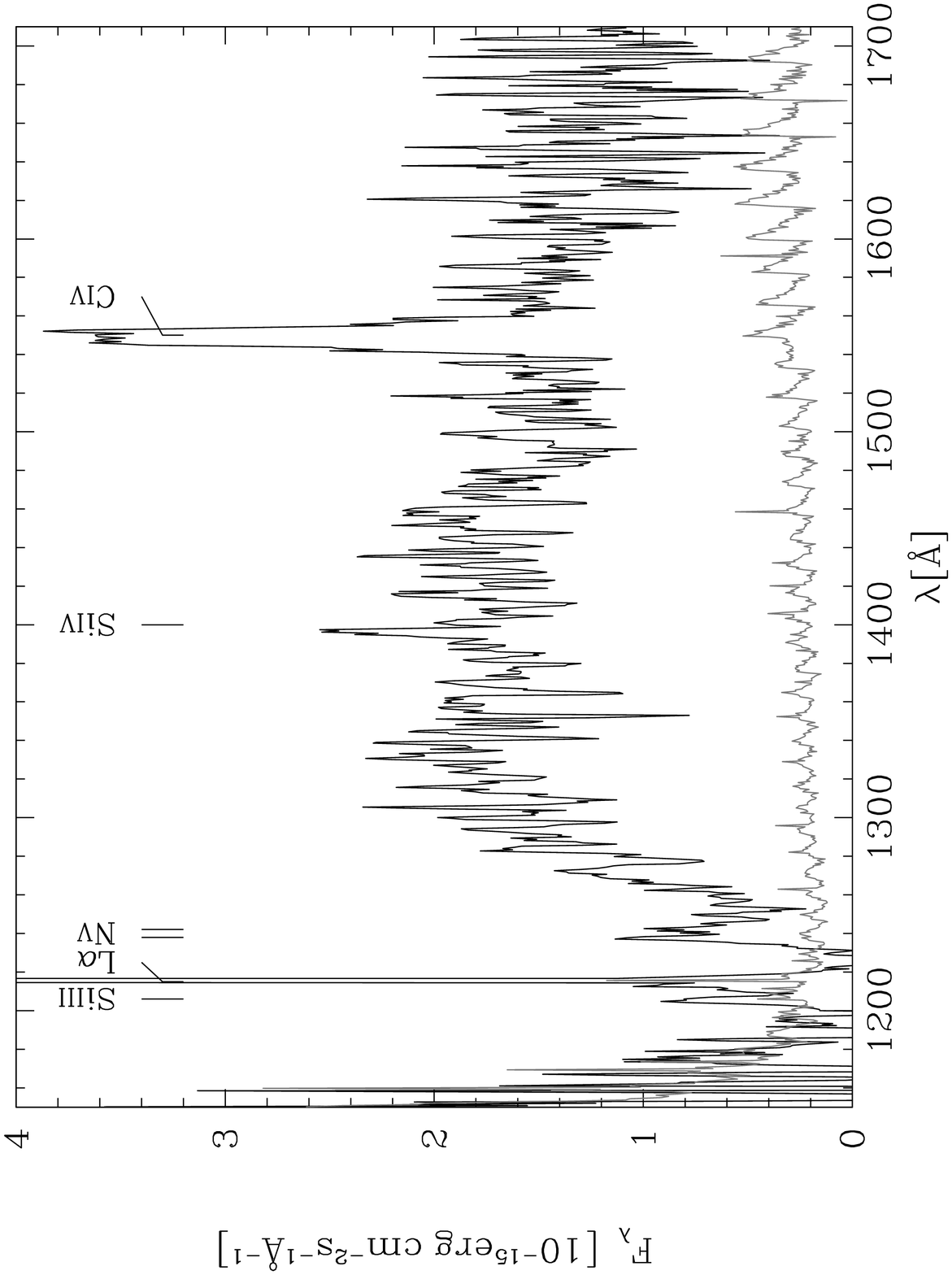,width=6in,angle=0}
\end{figure}

\begin{figure}
\psfig{figure=fig2a.ps,width=6in,angle=0}
\end{figure}

\begin{figure}
\psfig{figure=fig2b.ps,width=6in,angle=0}
\end{figure}

\begin{figure}
\psfig{figure=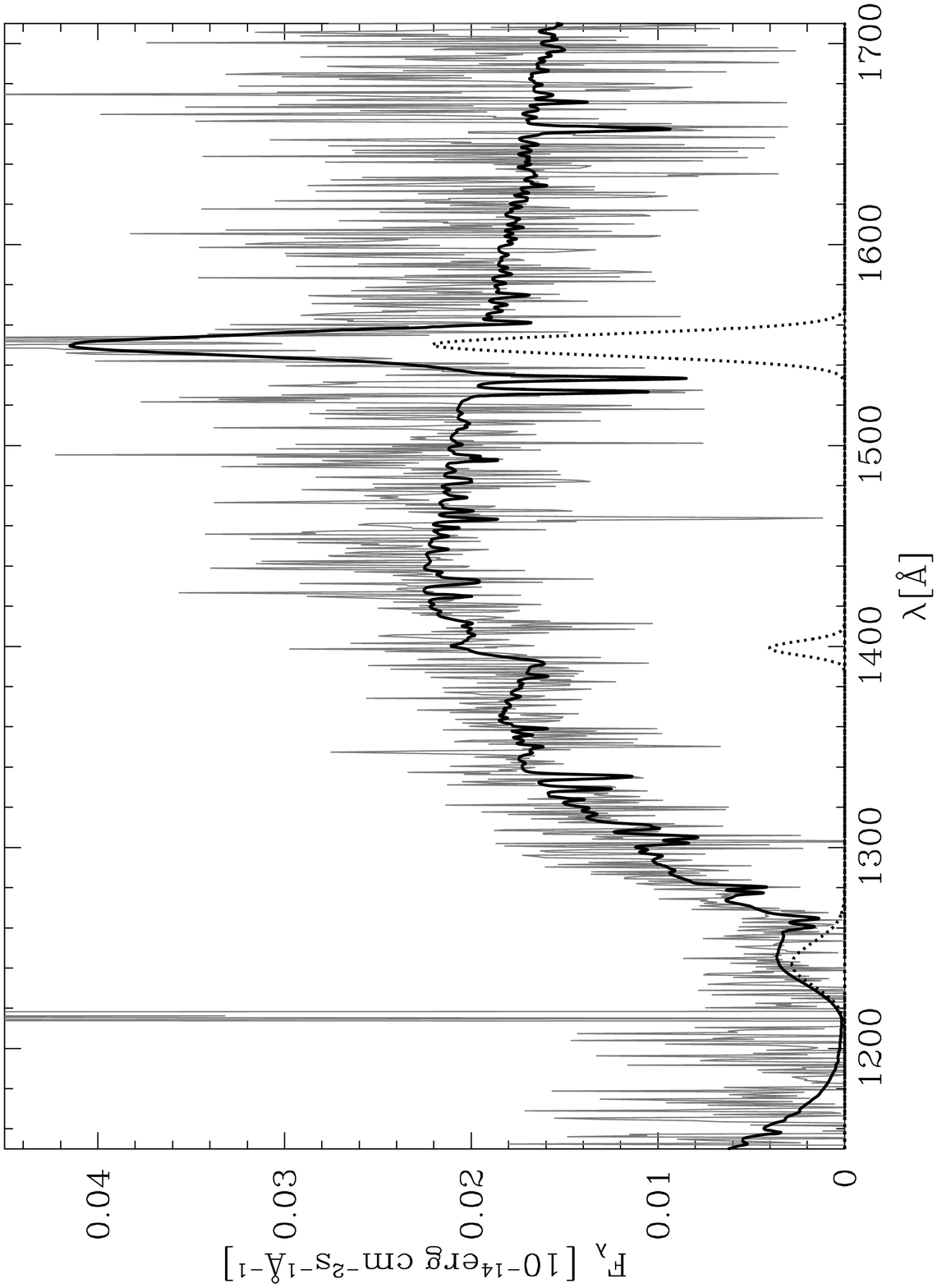,width=6in,angle=0}
\end{figure}

\begin{figure}
\psfig{figure=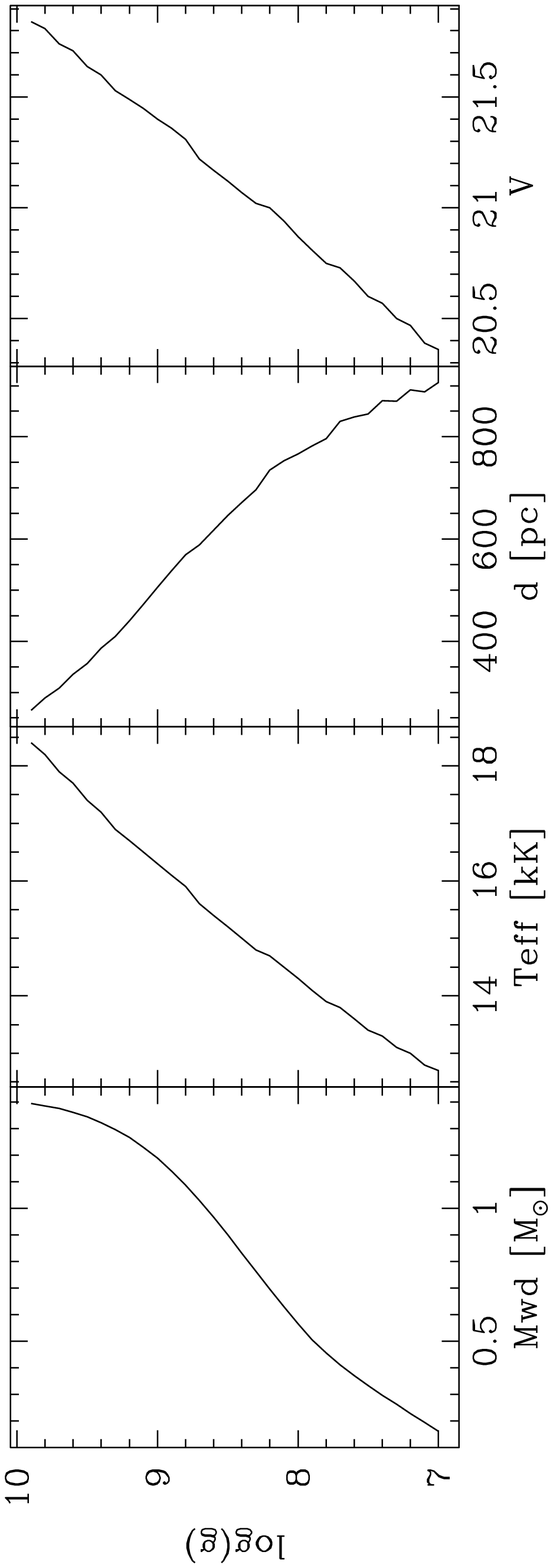,width=6in,angle=0}
\end{figure}

\begin{figure}
\psfig{figure=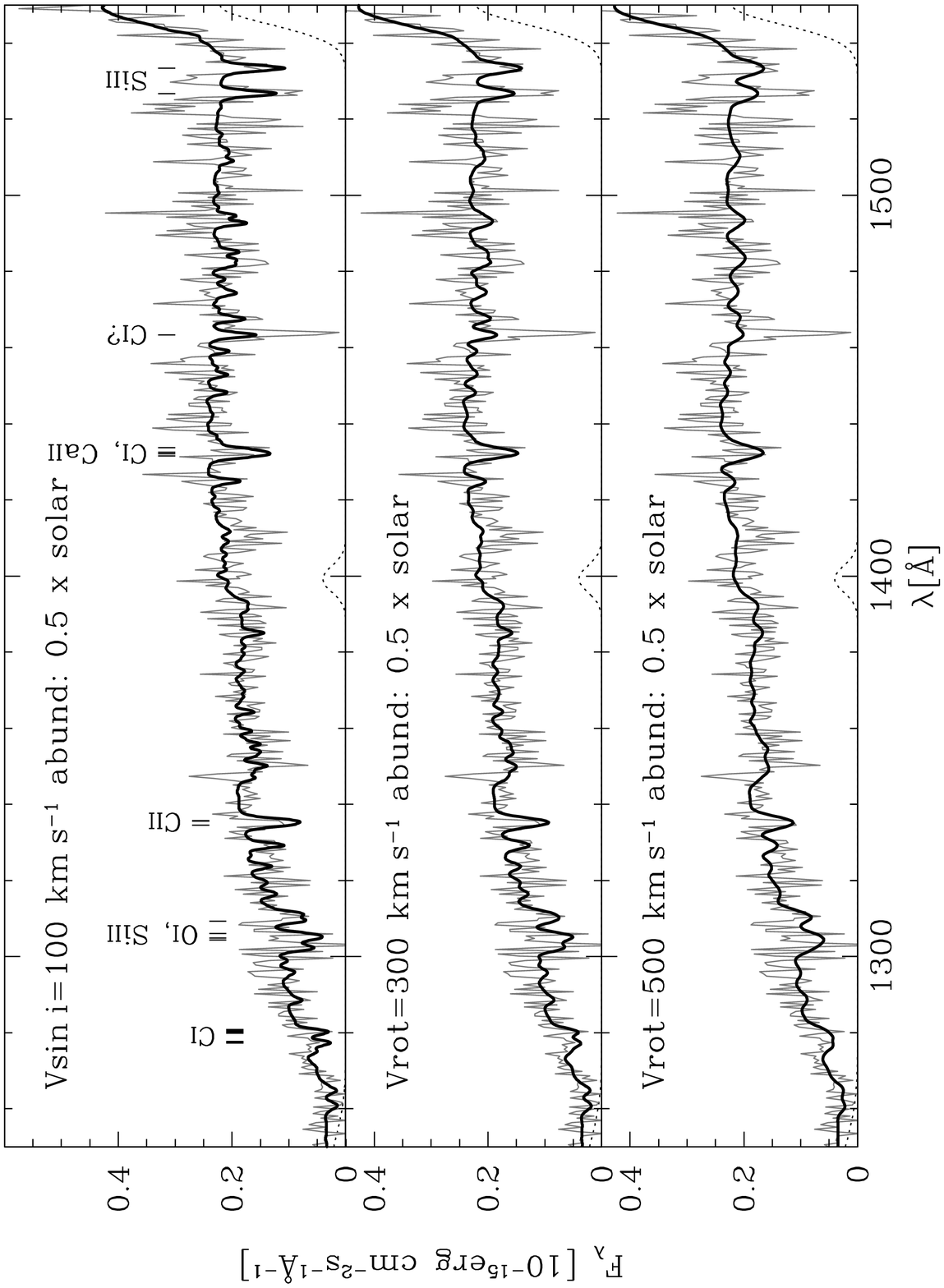,width=6in,angle=0}
\end{figure}

\begin{figure}
\psfig{figure=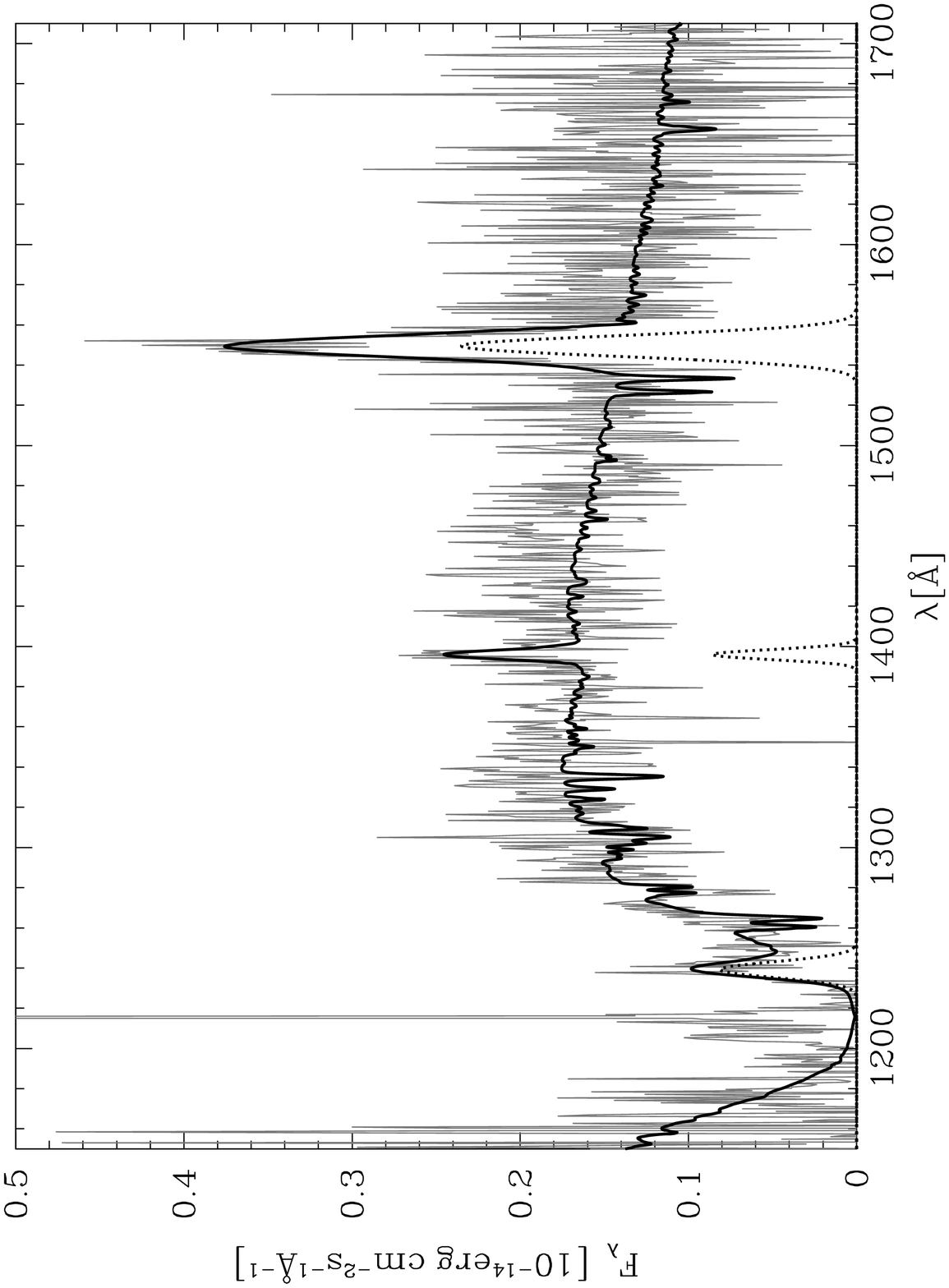,width=6in,angle=0}
\end{figure}

\begin{figure}
\psfig{figure=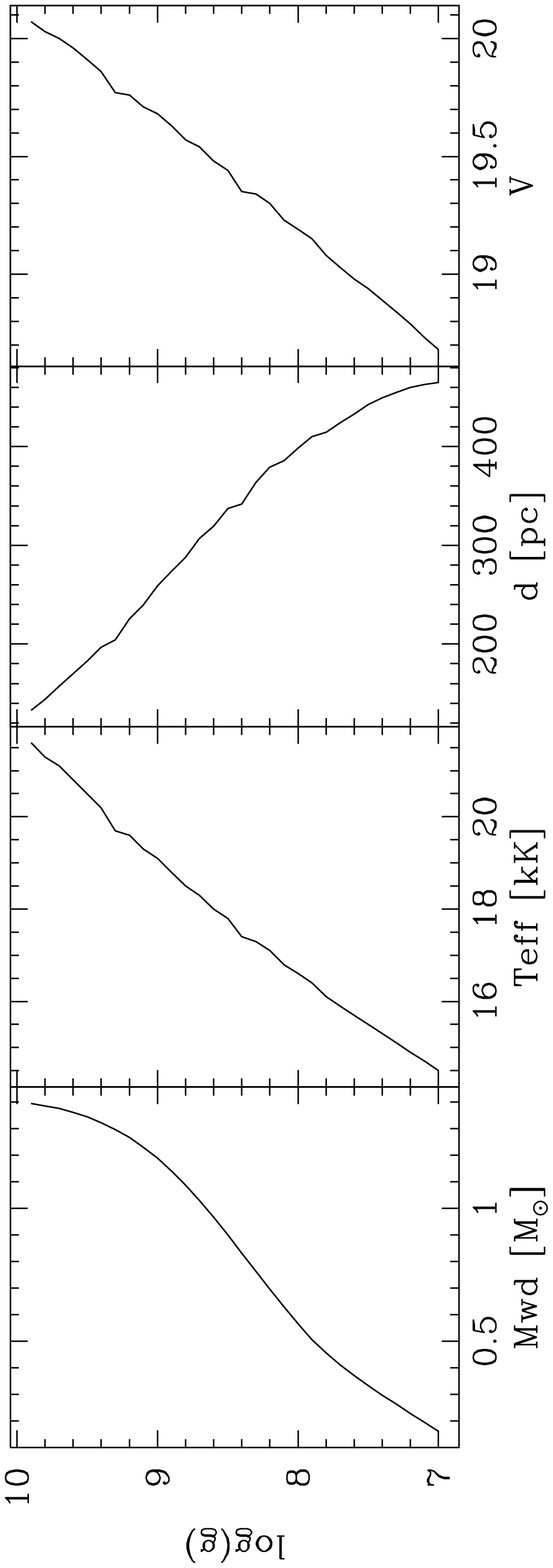,width=6in,angle=0}
\end{figure}

\begin{figure}
\psfig{figure=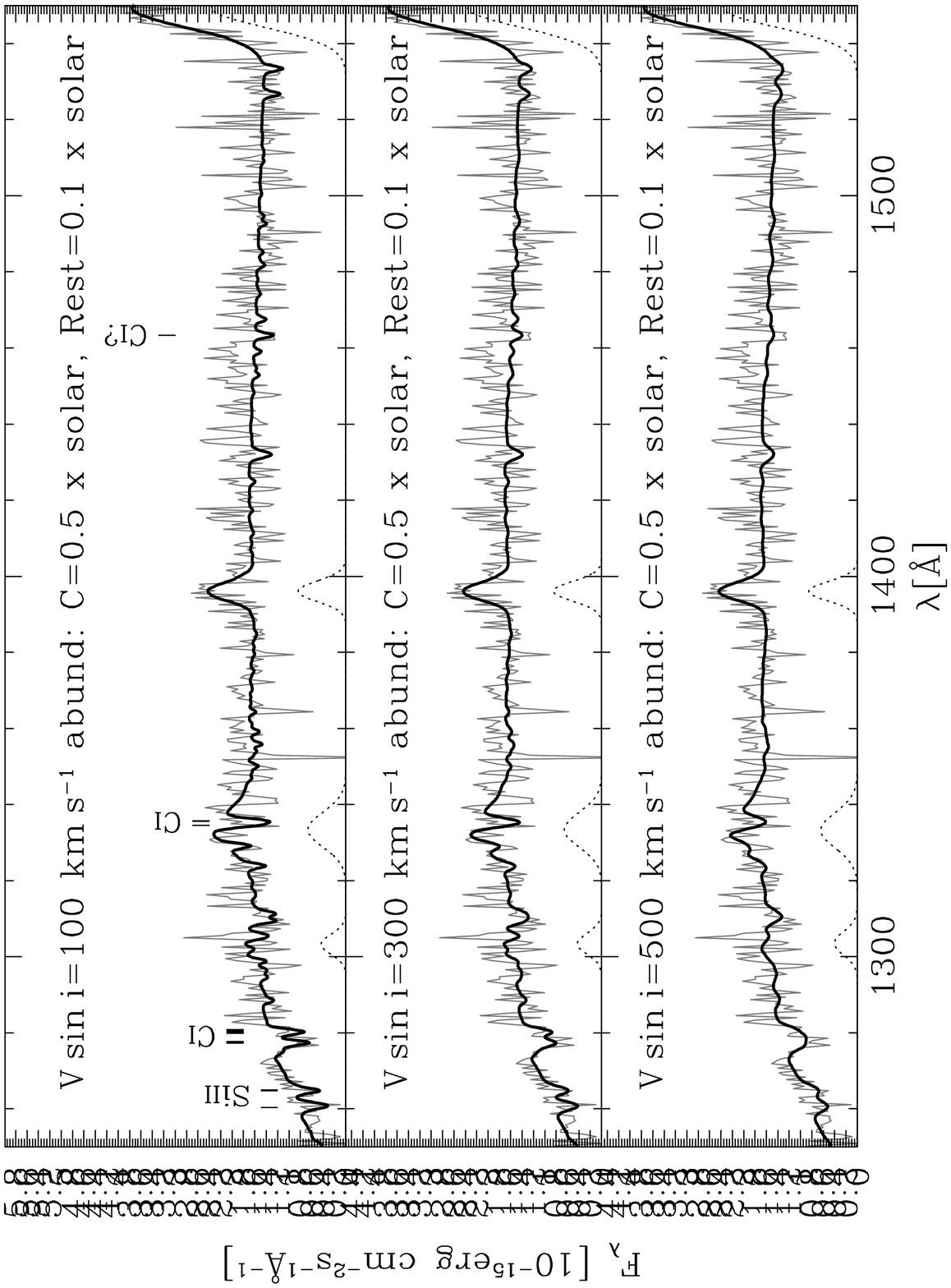,width=6in,angle=0}
\end{figure}

\begin{figure}
\psfig{figure=fig9.ps,width=6in,angle=0}
\end{figure}


\begin{figure}
\psfig{figure=fig10.ps,width=6in,angle=0}
\end{figure}


\end{document}